# Top-Down Design of a Low-Power Multi-Channel 2.5-Gbit/s/Channel Gated Oscillator Clock-Recovery Circuit

Paul Muller[1], Armin Tajalli[2], Mojtaba Atarodi[2], Yusuf Leblebici[1]

[1] Ecole Polytechnique Fédérale de Lausanne (EPFL)
Microelectronic Systems Laboratory
CH-1015 Lausanne, Switzerland

[2] Sharif University of Technology
Sharif Integrated Circuit and Systems Group
Teheran, Iran

### Abstract

*We present a complete top-down design of a low-power multi-channel clock recovery circuit based on gated current-controlled oscillators. The flow includes several tools and methods used to specify block constraints, to design and verify the topology down to the transistor level, as well as to achieve a power consumption as low as 5mW/Gbit/s. Statistical simulation is used to estimate the achievable bit error rate in presence of phase and frequency errors and to prove the feasibility of the concept. VHDL modeling provides extensive verification of the topology. Thermal noise modeling based on well-known concepts delivers design parameters for the device sizing and biasing. We present two practical examples of possible design improvements analyzed and implemented with this methodology.*

## 1. Introduction

While processor clock frequencies and throughput increase with each new technology generation, the lack of I/O bandwidth in microprocessors is an increasing limitation of the overall communication performance of computers. Short-distance communication interfaces like computer buses and LAN systems must support higher data rates to keep the pace with the evolution of processor speed. Parallel buses send a clock signal on a separate path to the receiver. Clock skew, due to unequal path length and termination impedance, is one limiting factor of the achievable data rate on printed circuit board buses. Crosstalk generated by the relatively large signal amplitude in parallel buses, as well as ringing due to impedance variations and mismatch are other limitations. Finally the high power drain of rail-to-rail output drivers for several tens of data lanes and the resulting possible ground bounce are other problems of parallel data communication interfaces.

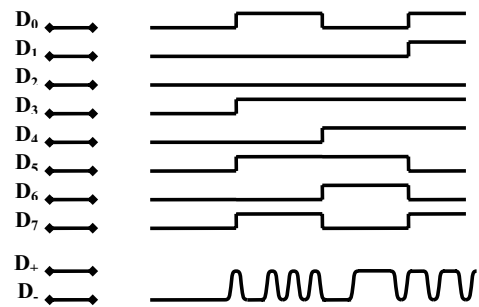

*Fig. 1: Parallel 8-bit bus versus serial communication with equivalent data rate*

Serial communications offer an important increase in data rates. Low-voltage differential signaling (LVDS) alleviates crosstalk and ground bounce, while point-to-point connections reduce impedance variations. Clock skew is not limiting the data rates in short-distance serial interfaces, as the clock is not transmitted as a separate signal. Indeed, the clock information is embedded in the data stream using 8bit/10bit encoding, which guarantees a high number of data transitions (Figure 1).

Short-distance optical links provide an even more robust solution to electro-magnetic coupling and bandwidth requirements. The presented clock and data recovery (CDR) solution combines with the already demonstrated pure silicon amplification front-end ([1], [2]) to a complete fiber-optic receiver. Mixed-signal modeling of such a system, as discussed in this paper, is a crucial step in the definition of the block-level specifications.

A multi-channel serial-link combines several serial data links side by side. Although such wide-band very-short reach multi-channel links are limited to server interconnects (e.g. InfiniBand™) today, they will certainly progress from board-to-board to chip-to-chip links, to be integrated on backplanes and microprocessor motherboards in the near future (Figure 2). Unlike in a parallel bus, the individual links are not, and do not need to be, fully synchronous.

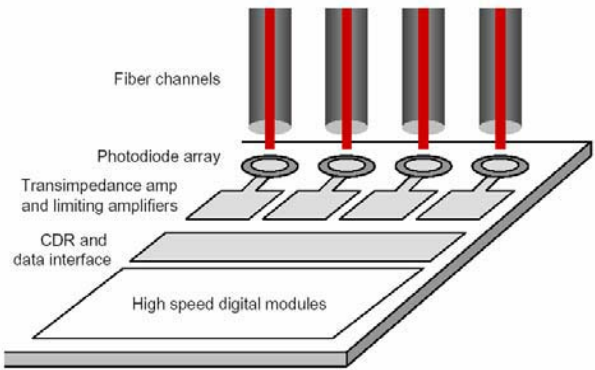

*Fig. 2: Conceptual block diagram of an integrated multi-channel photo-receiver array for data communication*

The presented design methodology allows us to design a low-power multi-channel clock and data recovery circuit with a power consumption lower than 5mW/Gbit/s. To achieve this low power consumption, we do not intend to use popular PLL, DLL or phase interpolation techniques, but to implement a gated oscillator topology [3], which is used relatively rarely.

Statistical simulations show that the gated oscillator approach is a viable solution in presence of frequency and phase variations. Then, the required power consumption is estimated based on the analysis of the phase noise of the oscillator. Behavioral VHDL simulations verify the time-domain behavior of the design and finally SPICE-level simulations validate the transistor-level design. As such, the presented design methodology demonstrates the feasibility of a top-down approach based on quantifiable system specifications, as opposed to classical bottom-up design.

## 2. System-level specifications

### 2.1. Definition of jitter tolerance

Because of the skew between the links in a multi-channel serial interface, individual synchronization of each channel is necessary. As the transmitter reference clock is used for all channels of a given transmitter, their data rates are identical. But each channel may exhibit a different delay, introducing skew between the channels. The CDR at the receive side must thus extract the phase of the incoming data of each channel.

The lowest bit error ratio (BER) is achieved when the data is sampled at the ideal sampling instant in between two data transitions, in most cases in the middle of the data eye (Figure 3). Timing jitter expresses the uncertainty of the sampling instant due to noise (random jitter) or systematic errors (deterministic jitter). In short-haul communications, the resynchronized data is transferred from the receive clock domain to the system clock domain through an elastic buffer (Figure 4), which resynchronizes the data with the system clock.

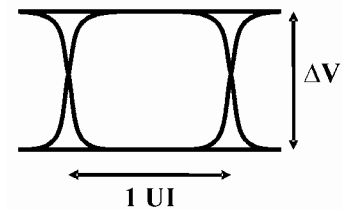

*Fig. 3: Data eye diagram with optimum sampling point*

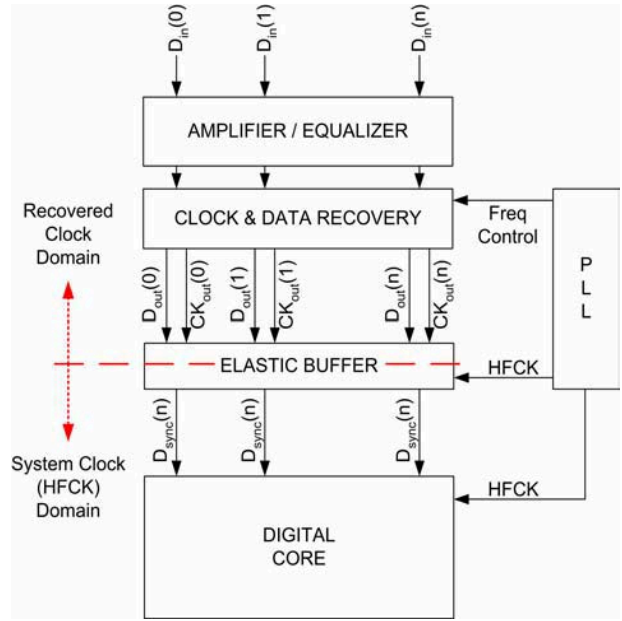

*Fig. 4: System view of digital core with serial I/O*

Tolerance to data jitter (JTOL) is usually tested by adding sinusoidal jitter at a given frequency to the data stream, which already includes channel jitter. The maximum jitter amplitude, function of jitter frequency, at which the CDR still operates at a given BER (typically $10^{-12}$), is called jitter tolerance (Figure 5). Jitter amplitude is typically represented in unit intervals (UI) either peak-peak or RMS, depending on the nature of jitter. A unit interval represents the bit period of the incoming data, i.e. 1UI = 400ps.

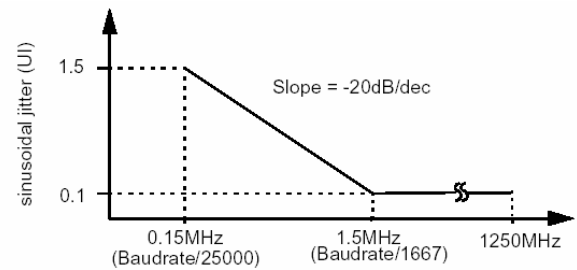

*Fig. 5: InfiniBand™ jitter tolerance specification[4]*

## 2.2. Gated oscillator topology

Multi-channel CDRs require high jitter performance and low area and power consumption at the same time. The gated current-controlled oscillator (GCCO) topology represents a good trade-off for this application. As shown in Figure 6, a shared PLL generates the local high-frequency clock (HFCK) from a low-frequency crystal oscillator clock (LFCK). The shared PLL is based on a high-order loop filter to get good jitter performance and a current-controlled oscillator (CCO). It also delivers a copy of its control current $I_C$ to the matched oscillators in each channel. Provided the CCOs are well matched, the clock frequencies of all channels (RXCK0..n) are identical to HFCK.

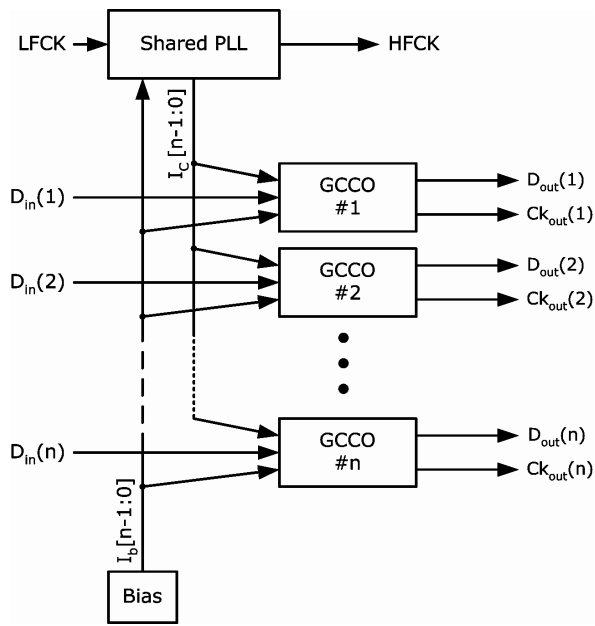

*Fig. 6: Multi-channel GCCO CDR*

At each data edge, an edge detection circuit based on a delay line and an XOR gate generates a synchronization signal EDET for the GCCO (Figure 7). For better isolation, the whole design is realized using fully-differential current-mode logic gates.

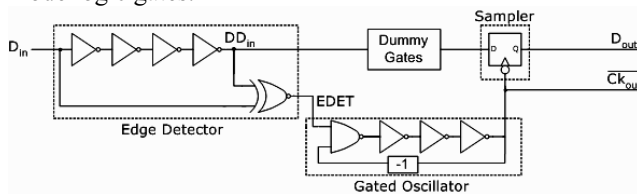

*Fig. 7: Gated oscillator with edge detection*

At an incoming data edge ($D_{IN}$), EDET goes low for a duration defined by the delay of the edge detector delay line (Figure 8). The first stage of the oscillator is frozen to a high state, which propagates through the oscillator and reaches the clock output after half a clock period (T/2).

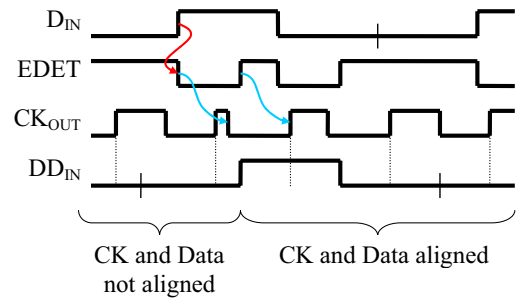

*Fig. 8: Timing diagram of GCCO*

At the rising edge of EDET, the oscillator is released and goes back to free oscillation at the frequency determined by its control current. As the data fed to the sampler ($DD_{IN}$) is taken at the output of the edge detector, the delay and jitter introduced by the delay line do not influence on the precision of the sampling. Parasitic delays coming from the XOR gate or the delay mismatch between both inputs of the NAND gate in the oscillator are compensated for by dummy gates. All delay cells in the delay line and the ring oscillator are built with identical current-mode logic two-input gates.

## 2.3. Frequency tolerance

Unlike in PLL-based or phase-interpolated clock recovery circuits, there may be a frequency difference between the gated oscillator in the receiver of a given channel and the incoming data stream. In practical applications, the data rate is specified to ±100ppm. The frequency tolerance (FTOL), defined as the maximum frequency difference at which the BER remains lower than $10^{-12}$ specifies the requirements on frequency stability of the gated oscillator design.

8bit/10bit encoding schemes used in short-distance communications reduce the effective data rate by 20%, but limit the number of consecutive identical digits (CID) to five. This is the worst case for accumulation of jitter and frequency error, to be taken into account in the analysis of JTOL and FTOL.

## 3. Behavioral modeling

### 3.1. Statistical model

Considering the gated CCO topology shown in Figure 7, we implemented a Matlab model to analyze JTOL and FTOL and the resulting BER with respect to incoming jitter, oscillator jitter and CID. In statistical models, the exact contributions of different types of timing jitter can be accurately combined. Deterministic jitter is modeled with a uniform probability density function (PDF), random jitter with a normal PDF and sinusoidal jitter leads to a sine wave histogram distribution.

The following simulations were all performed with the jitter specifications given in Table 1.

*Table 1: Jitter specifications for simulations*

| Jitter Type | Units | Value |
|---|---|---|
| Deterministic (DJ) | $UI_{PP}$ | 0.4 |
| Random (RJ) | $UI_{RMS}$ | 0.021 (0.3 $UI_{PP}$) |
| Sinusoidal (SJ) | $UI_{PP}$ | swept |
| Oscillator (CKJ) | $UI_{RMS}$ | 0.01 |

Figure 9 shows the achievable bit error rate when applying additional sinusoidal jitter. The targeted bit error rate of $10^{-12}$ is much above the specifications of Figure 5, especially for low-frequency jitter.

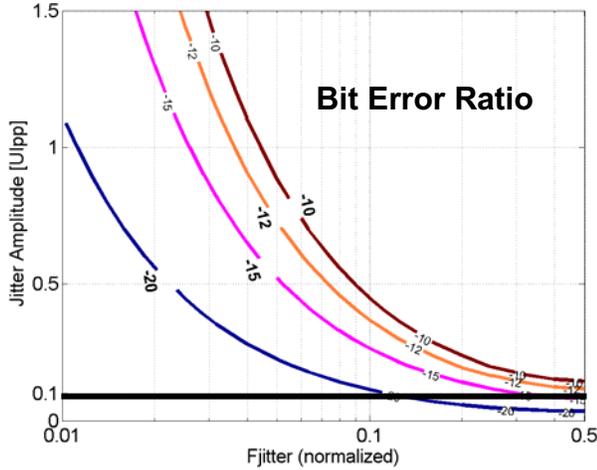

*Fig. 9: BER as a function of sinusoidal jitter frequency (normalized to data rate) and amplitude*

When frequency offset is present, the accumulated frequency difference over several CID is harmful to the performance of the gated-oscillator CDR (Figure 10).

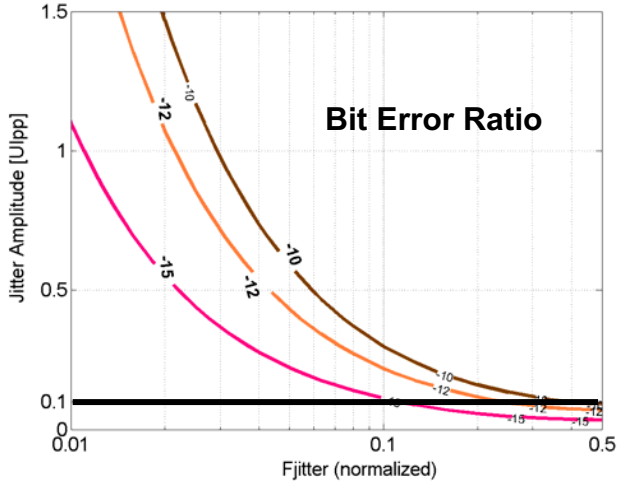

*Fig. 10: BER with frequency offset of 1%*

For jitter frequencies close to the data rate, the estimated jitter tolerance for a BER of $10^{-12}$ drops below the tolerance mask. This shows that there is very little design margin in oscillation frequency and oscillator phase noise.

### 3.2. Phase noise estimation

Frequency stability, already discussed, and timing jitter are the two most important specifications of the oscillator in the GCCO topology. Timing jitter of ring oscillators, or its frequency domain analogy phase noise, have been extensively studied [5]-[7].

$$\kappa_{min} = \sqrt{\frac{8kT}{3\eta I_{SS}}\left(\frac{1}{\Delta V/\gamma} + \frac{1}{R_L I_{SS}}\right)} \quad (eq.1)$$

Equation 1 (Hajimiri) allows us to derive the phase noise - power consumption trade-off of the ring oscillator, shown in Figure 11, in comparison with a variation of McNeill's formula. $I_{SS}$ is the bias current of one oscillator stage, $R_L$ its load resistance, $\Delta V$ the signal swing, $\gamma$ the noise factor of the active devices and $\eta$ indicates the relationship between rise-time and cell delay.

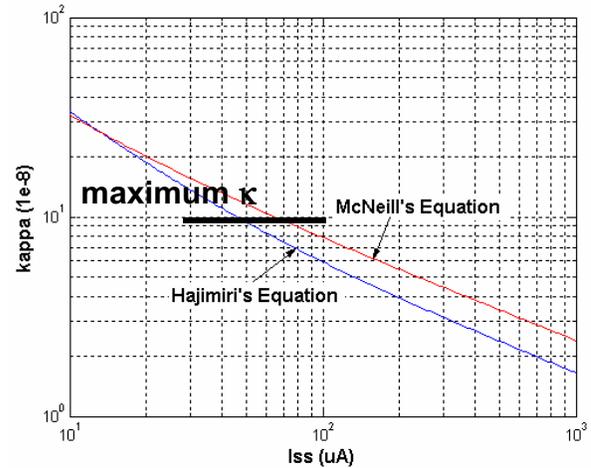

*Fig. 11: Phase noise–power consumption trade-off*

The oscillator bias currents and derived device dimensions are chosen based on this graph. The respective standard deviation for the sampling clock is $0.01 UI_{RMS}$ for CID = 5.

### 3.3. VHDL model

All system components have been modeled in VHDL to perform behavioral verification in the time domain. The statistical model previously discussed gives a good estimate of achievable performance, but it would be difficult to introduce temporal behavior and non-ideal gate delays in the statistical models. The behavioral model is close to the physical implementation, as shown in Figure 12, and considers most contributions to jitter and static phase error.

Unlike the transistor-level simulation, the CPU requirements are sufficiently low to run parametric simulations on frequency offset and jitter contributions (10μs simulated in ~ 45min). The gate-level VHDL model, taking into account jitter components and delay non-idealities, can thus be used for extensive verification of the CDR topology.

VHDL and Matlab simulations are based on the same specifications in terms of jitter and frequency tolerance. Deterministic, random and sinusoidal jitter are applied to the incoming data stream. The phase noise of each cell of the delay line and the GCCO is independently calculated as a random delay variation. Amplitude noise is neglected. This assumption is commonly accepted for clock recovery circuits, as the pre-amplification in the system delivers binary signals.

```
entity cdr_gcco is
 generic (
   cdr_gcco_k: real;        -- CCO gain [Hz/A]
   cdr_gcco_fc: real;       -- Free-running frequency [Hz]
   cdr_gcco_cc0: voltage;   -- Control current mid-point [C]
   cdr_gcco_jit_sigma: real);  -- defined as a ratio, e.g 1%->0.01
 port (
   […]);
end entity cdr_gcco;

architecture bhv of cdr_gcco is
 […]
begin  -- ring_osc
 calc_delay0: process
 begin  -- process calc_delay
   awgn(seed1, seed2, mean, sigma, jitter);  -- gaussian random gen. [8]
   delay0 <= 1 ps * 1.0e12/
     (8.0*(cdr_gcco_fc+cdr_gcco_k*(cctrl-cdr_gcco_cc0))) * (1.0+jitter);
   wait for delay0;
 end process calc_delay0;
 […] -- calculation of the three remaining delays
 vinv1(0) <= transport (vinv4(0) and cdr_gcco_trig(0)) and
     (cdr_gvco_enable and cdr_gcco_nreset) after delay0;
 vinv1(1) <= transport ((vinv4(0) nand cdr_gcco_trig(0)) and
     (cdr_gvco_enable and cdr_gcco_nreset)) after delay0;
 vinv2 <= transport not(vinv1) after delay1;
 vinv3 <= transport not(vinv2) after delay2;
 vinv4 <= transport not(vinv3) after delay3;
 cdr_gcco_ckout <= not(vinv4);
end bhv;
```

*Fig. 12: VHDL code of gated CCO*

The contribution of the VHDL model to design verification will be illustrated by two concrete issues raised and solved with this approach.

*a. Delay in the edge detector*

The behavioral models include gate delays and timing jitter in the ring oscillator components. The current-mode logic cells used in this design exhibit different input to output delays for the different inputs, due to the stacked nature of the design. These different delays must be taken into account in this kind of simulation.

Although the chosen topology should be insensitive to the exact delay of the delay line in the edge detector, VHDL simulations raised concerns about this hypothesis. Figure 13 shows that if the rising edge of EDET occurs before the forth stage of the oscillator ($CK_{OUT}$) has gone low, the oscillator will not be synchronized with the respective data edge.

If the oscillator frequency is close to the data rate, this can happen for tens or more consecutive data edges, leading to poor jitter tolerance. Detailed analysis showed that reliable operation is guaranteed for $T/2 < \tau < T$.

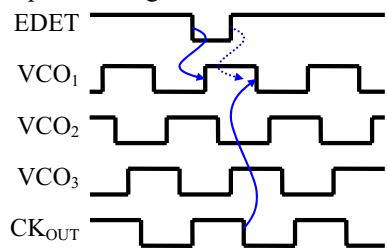

*Fig. 13: Problem situation for $\tau \leq T/2$*

*b. Optimum sampling point*

An even more interesting consideration in the behavioral simulations is the selection of the optimum sampling instant. As the oscillator is triggered by each incoming data edge, the left data edge in the eye diagram has a narrow distribution, while the right edge suffers of accumulated jitter and frequency error over several CID (Figure 14).

In order to plot the output data eye diagram as it occurs at the sampler input, an *eye generator* block in VHDL has been inserted. Unlike eye diagram features in conventional tools, it does not use fixed time sampling intervals, but aligns the data on the rising edge of the sampling clock. The aligned data is send to a text file, which can be easily read into Matlab to plot the eye diagram shown below.

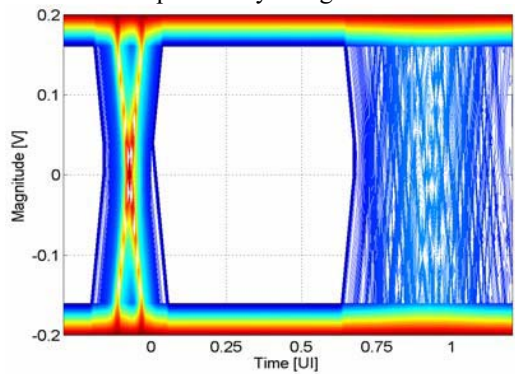

*Fig. 14: 25k cycles PRBS7 eye diagram simulated in VHDL with: CCO frequency = 2.375GHz sin. jitter amp = 0.10UI$_{pp}$, freq = 250 MHz*

In order to increase the BER, the sampling instant has been shifted by one eighth of the clock period, using the inverted output of the third inverter stage as recovered clock (Figure 15).

As the whole CDR is designed using fully differential logic gates for proper high speed operation, the change of sign of the clock signal compared to the topology in Figure 7 is taken care of by inverting the differential output signal. This does not require any additional logic gates, which would introduce critical delay in the clock path.

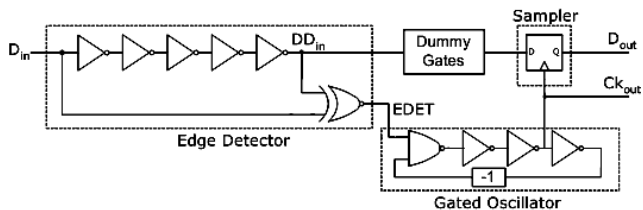

*Fig. 15: Modified GCCO topology*

The simulation results in Figure 16 show an obvious improvement in timing margin on the right data edge, i.e. the eye opening is almost symmetrical around UI/2. Note that in both cases, a standard pseudo-random bit sequence (PRBS7) was applied, which exhibits more consecutive identical digits than an 8bit/10bit encoded stream.

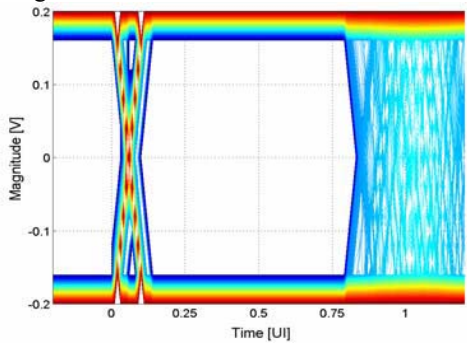

*Fig. 16: Eye diagram with improved oscillator output (same conditions)*

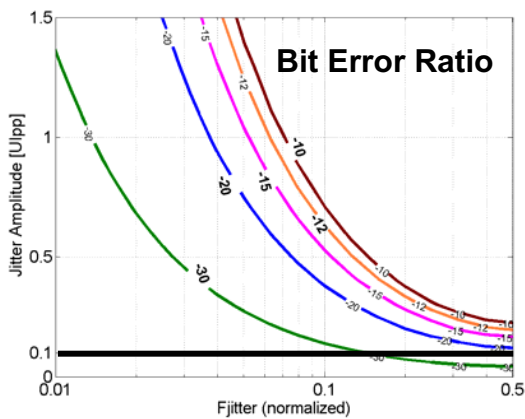

*Fig. 17: BER estimation with frequency error of 1% with improved sampling point*

Statistical bit error rate estimations also show improved results when taking into account the shift of the sampling point, based on the conditions of Figure 10 (Figure 17). While the topology directly improves the horizontal eye opening as shown above, this sampling point may however increase the probability of erroneous sampling of the next bit due to frequency offset, not considered in Figure 17.

## 4. Transistor-level design

Based on the high-level simulations and phase noise estimations previously discussed, we designed the gated CCO clock and data recovery circuit in a 0.18µm CMOS digital process from UMC.

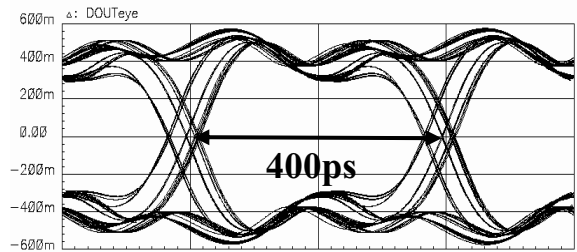

*Fig. 18: Eye diagram from transistor-level simulation (typical case, no jitter applied)*

While detailed discussion of the transistor-level design is beyond the scope of this paper, Figure 18 shows the output eye diagram obtained in typical case simulation. The layout of the test chip is to be completed by the time of publication. Experimental results will be presented in a follow-up paper.

## 5. Conclusion

We presented a design flow to achieve a low-power multi-channel clock and data recovery circuit. The flow allowed us to have a good estimate of the achievable bit error rate of this topology through statistical simulations. Using behavioral VHDL modeling, we were able to improve the bit error rate of this solution by introducing minor changes in the topology. These changes were extensively verified in various jitter and frequency offset configurations. The design flow presented here serves as a convincing demonstration that a complete top-down approach can be implemented in the design of demanding high-speed analog ICs.

This research has been supported in part by the Swiss National Science Foundation Grant 200021-100625.